\documentclass{article}
\usepackage{arxiv}
\usepackage[utf8]{inputenc} % allow utf-8 input
\usepackage[T1]{fontenc}    % use 8-bit T1 fonts
\usepackage{hyperref}       % hyperlinks
\usepackage{url}            % simple URL typesetting
\usepackage{booktabs}       % professional-quality tables
\usepackage{amsfonts}       % blackboard math symbols
\usepackage{amsmath}        % math environments
\usepackage{nicefrac}       % compact symbols for 1/2, etc.
\usepackage{microtype}      % microtypography
\usepackage{lipsum}
\usepackage{graphicx}
\graphicspath{ {./images/} }

\title{Mass-Conserving Physics-Informed Neural Networks for the One-Dimensional Advection-Diffusion Equation}

\author{
  Eszra Forenita Sigalingging \\
  Department of Physics, Faculty of Mathematics and Natural Sciences\\
  Universitas Padjadjaran\\
  Sumedang, Indonesia \\
  \texttt{eszra22001@mail.unpad.ac.id} \\
  \And
  Liu Kin Men \\
  Department of Physics, Faculty of Mathematics and Natural Sciences\\
  Universitas Padjadjaran\\
  Sumedang, Indonesia \\
  \texttt{liu.kin.men@unpad.ac.id} \\
  \And
  Setianto \\
  Department of Physics, Faculty of Mathematics and Natural Sciences\\
  Universitas Padjadjaran\\
  Sumedang, Indonesia \\
  \texttt{setianto@phys.unpad.ac.id} \\
  \And
  Ferry Faizal \\
  Department of Physics, Faculty of Mathematics and Natural Sciences\\
  Universitas Padjadjaran\\
  Sumedang, Indonesia \\
  \texttt{ferry.faizal@unpad.ac.id} \\
}

\begin{document}
\maketitle

\begin{abstract}
The advection-diffusion equation is a fundamental model of transport phenomena in which mass conservation is an essential physical constraint. While classical schemes such as Crank-Nicolson preserve this property by construction, Physics-Informed Neural Networks (PINNs) enforce only the local residual of the governing PDE and are therefore not guaranteed to conserve global quantities such as mass over long integration horizons. In this work, we examine the extent of this limitation for the periodic one-dimensional advection-diffusion equation and evaluate a Mass-Penalty PINN that augments the standard PINN loss with a soft mass-conservation constraint. We compare the performance of Vanilla PINN, Mass-Penalty PINN, and the Crank-Nicolson scheme across a range of Peclet numbers spanning diffusion-dominated to advection-dominated regimes, and over two simulation horizons representing short-term and long-term dynamics. The results show that, for short-term simulations, the Mass-Penalty PINN does not always provide a consistent improvement in accuracy. However, for long-term simulations, the Mass-Penalty PINN reduces the relative $L_2$ error and mass conservation error by factors of approximately 9-67 and 15-215, respectively, compared with the Vanilla PINN, across the tested Peclet numbers. Further analysis reveals that the accuracy degradation observed in Vanilla PINN is predominantly caused by the accumulation of mass drift over time. These results demonstrate that incorporating a soft mass-conservation constraint substantially improves the long-term reliability of PINN for conservative transport problems, particularly in mitigating mass drift over extended simulation horizons.
\end{abstract}


\section{Introduction}

Advection-Diffusion Equation (ADE) is one of the fundamental partial differential equations (PDE) used in transport physics modeling. It combines the effect of advection, which transports a physical quantity through the motion of a fluid or medium, and diffusion which redistributes the quantity due to concentration gradients. This equation is widely applied in science and engineering fields, such as water quality modeling \cite{Mugheri2023EFFICIENTENA}, contaminant transport \cite{Aghdam2020ACAA}, and air quality prediction \cite{hettige2024airphynetharnessingphysicsguidedneural}. In many practical applications, the advection-diffusion equation should satisfy the conservation laws. For a system without internal sources or sinks, the total mass or energy contained within the domain should remain constant over time. Classical numerical methods, particularly those rooted in the finite volume method (FVM) \cite{LeVeque_2002,eymard}, are designed to preserve these conservation properties at the discrete level. Among finite difference methods, the Crank-Nicolson (CN) method \cite{crank1947practical} is highly regarded for its unconditional stability and second-order accuracy in both space and time, making it a robust baseline for solving parabolic PDEs like ADE \cite{Hundsdorfer2003}. However, classical grid-based methods often struggle with high-dimensional spaces, complex geometries, and inverse parameter estimation problems.

Physics-Informed Neural Networks (PINN) introduced in 2019 by Raissi et al. \cite{RAISSI2019686} represent a mesh-free approach to solving partial differential equations (PDEs) by embedding governing laws directly into the neural network loss function. By leveraging automatic differentiation (AD) \cite{baydin}, PINN can evaluate continuous, differentiable solutions at arbitrary points, making them highly attractive for complex geometries and data-sparse inverse problems. The broader field of Scientific Machine Learning has rapidly adopted PINN, as evidenced by comprehensive reviews highlighting their versatility \cite{Karniadakis2021,Cuomo2022,cai2021physics}. Despite their elegance, standard PINN suffer from critical limitations when applied to advection-dominated or strictly conservative systems.

Recent studies have characterized the limitations of PINN in such regimes. Krishnapriyan et al. \cite{Krishnapriyan} demonstrated that PINN can struggle to learn the correct solution for convection-dominated flows, often converging to trivial or highly inaccurate states. Furthermore, Mamud et al. \cite{Mamud2024} demonstrated that Vanilla PINN achieves accurate solutions for groundwater flow but fails to maintain mass balance, particularly over long simulation times. This mass drift phenomenon causes solutions to converge toward an incorrect average concentration level, leading to significant accuracy degradation in long-time simulations. The root causes of these failures are often attributed to gradient pathologies \cite{wang2021understand} and the unequal convergence rates of different loss terms, as explained through the Neural Tangent Kernel (NTK) framework \cite{wang2022when}.

To address these limitations, researchers have developed structure-preserving and conservative PINN variants. Approaches include conservative PINN (cPINN) utilizing domain decomposition \cite{djagtap2020,SHUKLA2021110683}, variational PINN (VPINN) based on weak formulations \cite{KHARAZMI2021113547}, and finite volume-inspired neural networks \cite{pradita,lichtle2025unfvsupervisedunsupervisedneural}. Other notable developments include integral PINN \cite{Wang2024}, Godunov-Riemann informed networks for hyperbolic laws \cite{patsatzis2025gorinns}, and mass-preserving spatio-temporal adaptive PINNs designed for highly nonlinear equations like Cahn-Hilliard \cite{huang2025mass}.

Despite these developments, the systematic evaluation of mass penalty effectiveness across different Peclet number regimes has not been thoroughly investigated for the advection-diffusion transport, and its influence on PINN performance and mass conservation behavior remains insufficiently characterized. This study addresses this gap by systematically evaluating three methods: Crank-Nicolson (CN), Vanilla PINN, and Mass-Penalty PINN against the periodic analytical solution across seven Peclet numbers (Pe = 0.01, 0.25, 0.5, 1, 5, 10, 20) at two simulation durations (T = 5\,s and T = 100\,s). The research focuses specifically on whether the mass penalty constraint effectively mitigates mass drift in long-time simulations, and on identifying Peclet regimes where each method performs reliably.

\section{Methodology}
\label{sec:methods}

\subsection{Governing Equation}

This study models the 1D advection-diffusion equation:
\begin{equation}
\frac{\partial c}{\partial t} + v \frac{\partial c}{\partial x} = D \frac{\partial^2 c}{\partial x^2}.
\label{eq:ade}
\end{equation}

The relative strength of advective to diffusive transport is characterized by the dimensionless Peclet number,
\begin{equation}
Pe = \frac{v L}{D},
\label{eq:peclet}
\end{equation}
where $L$ is a characteristic length scale of the domain. Diffusion dominates for $Pe \ll 1$, advection dominates for $Pe \gg 1$, and both mechanisms contribute comparably near $Pe \approx 1$.

\subsection{Initial and Boundary Conditions}

The initial condition is a Gaussian pulse centered at $x_0$ within a periodic domain $[0, L]$,
\begin{equation}
c(x,0) = \exp\left(-\frac{(x-x_0)^2}{4D\sigma^2}\right),
\label{eq:ic}
\end{equation}
with initial width parameter $\sigma$. Periodic boundary conditions,
\begin{equation}
c(0,t) = c(L,t), \qquad \frac{\partial c}{\partial x}(0,t) = \frac{\partial c}{\partial x}(L,t),
\label{eq:bc}
\end{equation}
are imposed so that no mass is allowed to leave the domain, isolating mass-conservation error as a property of the solution method rather than of the boundary treatment.

\subsection{Analytical Reference Solution}

Analytical solutions utilizing the Method of Images (periodized Gaussian kernels) were employed to objectively evaluate the PINN approaches:
\begin{equation}
c(x,t) = \sum_{m=-\infty}^{\infty} \frac{A\sigma}{\sqrt{\sigma^2 + 2Dt}} \exp\left(-\frac{(x - x_0 - vt + mL)^2}{2(\sigma^2 + 2Dt)}\right),
\label{eq:analytic}
\end{equation}
where $A$ is the amplitude of the initial pulse. Evaluating the total mass over the domain via the standard Gaussian integral shows that
\begin{equation}
M(t) = A\sigma\sqrt{2\pi},
\label{eq:mass_analytic}
\end{equation}
which is independent of both time and the advection velocity, confirming that the analytical solution is exactly mass-conserving and can therefore serve as ground truth for evaluating conservation error in the numerical schemes.

Additionally, high-resolution Crank-Nicolson solutions were generated to serve as an established numerical baseline. As a numerical baseline, Eq.~\ref{eq:ade} is discretized with the Crank-Nicolson scheme, which averages the spatial derivative over two successive time levels. The resulting discrete update takes the tridiagonal form
\begin{equation}
-\alpha\, c_{i-1}^{n+1} + (1+2\beta)\, c_i^{n+1} - \alpha\, c_{i+1}^{n+1} = \alpha\, c_{i-1}^{n} + (1-2\beta)\, c_i^{n} + \alpha\, c_{i+1}^{n},
\label{eq:cn}
\end{equation}
with $\alpha = v\Delta t / (4\Delta x)$ and $\beta = D \Delta t / (2\Delta x^2)$, solved at each time step via the Thomas algorithm.

A grid-independence study is performed over $N_x \in \{250, 500, 1000, 1500, 2000\}$, and confirmed that spatial resolution of $N_x = 1000$ guarantees an $L_2$ error below $10^{-6}$ over all tested Peclet numbers, establishing it as a robust numerical benchmark.

\begin{figure}[htbp]
\centering
\includegraphics[width=0.35\textwidth]{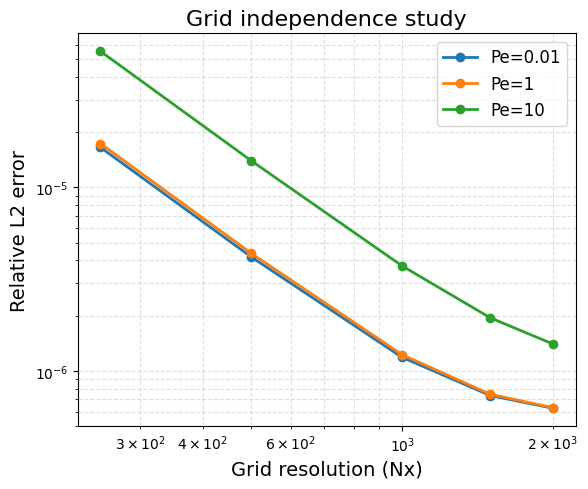}
\caption{Grid independence study for the Crank-Nicolson reference solution.}
\label{fig:grid_independence}
\end{figure}

\subsection{Physics-Informed Neural Network Formulation}

The concentration field is approximated by a fully connected neural network $c_\theta(x,t)$ with parameters $\theta$ (weights and biases), using the hyperbolic tangent activation function to guarantee smooth, infinitely differentiable outputs suitable for automatic differentiation \cite{Lu2019}. Training minimizes a composite loss function evaluated at three sets of collocation points sampled via Latin Hypercube Sampling: interior points enforcing the PDE residual, initial-time points enforcing Eq.~\ref{eq:ic}, and boundary points enforcing Eq.~\ref{eq:bc}.

\begin{figure}[htbp]
\centering
\includegraphics[width=0.8\textwidth]{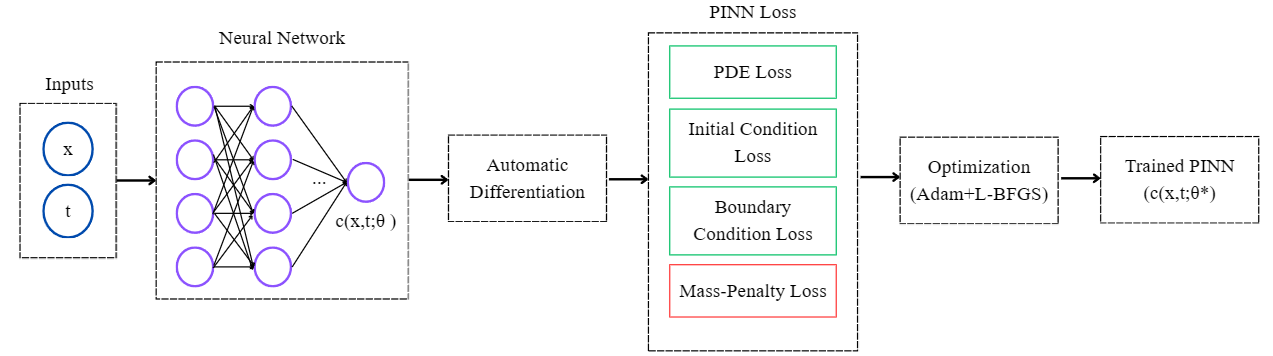}
\caption{PINN scheme implementation with mass penalty constraint}
\label{fig:pinn}
\end{figure}

The loss function for the Vanilla PINN minimizes the mean squared error of the PDE residual ($\mathcal{L}_{PDE}$), initial condition ($\mathcal{L}_{IC}$), and boundary condition ($\mathcal{L}_{BC}$). For the Mass-Penalty PINN, an additional soft constraint is integrated into the total loss following the mass-preserving strategy proposed by \cite{huang2025mass} in the context of the Cahn-Hilliard equation:
\begin{equation}
\mathcal{L}_m(\theta) = \frac{1}{N_t}\sum_{i=1}^{N_t} \left| m(c_\theta, t_i) - m(c_0) \right|^2,
\label{eq:mass_loss}
\end{equation}
\begin{equation}
m(c_\theta, t_i) = \int_{\Omega} c_\theta(x,t_i)\, dx \approx \sum_{j=1}^{N_x} \omega_j\, c_\theta(x_j^q, t_i),
\label{eq:mass_quad}
\end{equation}
where $m(c_0)$ is the initial analytical mass. The integral is computed using Gaussian Quadrature.

The total loss function becomes:
\begin{equation}
\mathcal{L}(\theta) = \lambda_{IC}\,\mathcal{L}_{IC}(\theta) + \lambda_{BC}\,\mathcal{L}_{BC}(\theta) + \lambda_{PDE}\,\mathcal{L}_{PDE}(\theta) + \lambda_m\,\mathcal{L}_m(\theta).
\label{eq:loss_masspenalty}
\end{equation}

\subsection{Training Procedure}

A preliminary hyperparameter tuning study was conducted to determine a robust PINN configuration before the main experiments \cite{Lu2019DeepXDEAD}. Hyperparameter optimization established the optimal penalty weights at $\lambda_{IC} = 100$, $\lambda_{BC} = 20$, $\lambda_{PDE} = 10$, and $\lambda_m = 10$. To prevent clustering in the loss landscape, the collocation points were sampled using Latin Hypercube Sampling (LHS), with 20,000 interior points, 4,000 initial condition points, and 4,000 boundary condition points. The network architecture was fixed at four hidden layers with 50 neurons per layer. Training utilized a two-stage strategy: 10,000 epochs of Adam optimization (learning rate $10^{-3}$) for rapid exploration, followed by 1,000 epochs of L-BFGS optimizer to refine gradient convergence. All computations were performed on an NVIDIA T4 GPU using PyTorch.

\subsection{Experimental Design}

Seven Peclet numbers ($Pe = 0.01, 0.25, 0.5, 1, 5, 10, 20$) are considered, spanning diffusion-dominated, transitional, and advection-dominated regimes. Each configuration is evaluated over two simulation horizons: a short-term horizon ($T=5$\,s) probing early spatio-temporal convergence, and a long-term horizon ($T=100$\,s) probing the accumulation of conservation error. Two metrics are used throughout. Mass conservation is quantified by the relative mass error,
\begin{equation}
E_{\mathrm{mass}} = \frac{|M(t) - M(0)|}{|M(0)|}
\label{eq:mass_error}
\end{equation}
and solution accuracy is quantified by the relative $L_2$ norm error against the analytical reference,
\begin{equation}
L_2 = \frac{\| c_{\mathrm{pred}} - c_{\mathrm{ref}} \|_2}{\| c_{\mathrm{ref}} \|_2}, \qquad \| c \|_2 = \left(\sum_{j=1}^{N} |c_j|^2\right)^{1/2}.
\label{eq:l2_error}
\end{equation}

\section{Numerical Results}
\label{sec:results}

\subsection{Short-Term Simulation Performance ($T = 5$\,s)}

\begin{figure}[htbp]
\centering
\includegraphics[width=0.7\textwidth]{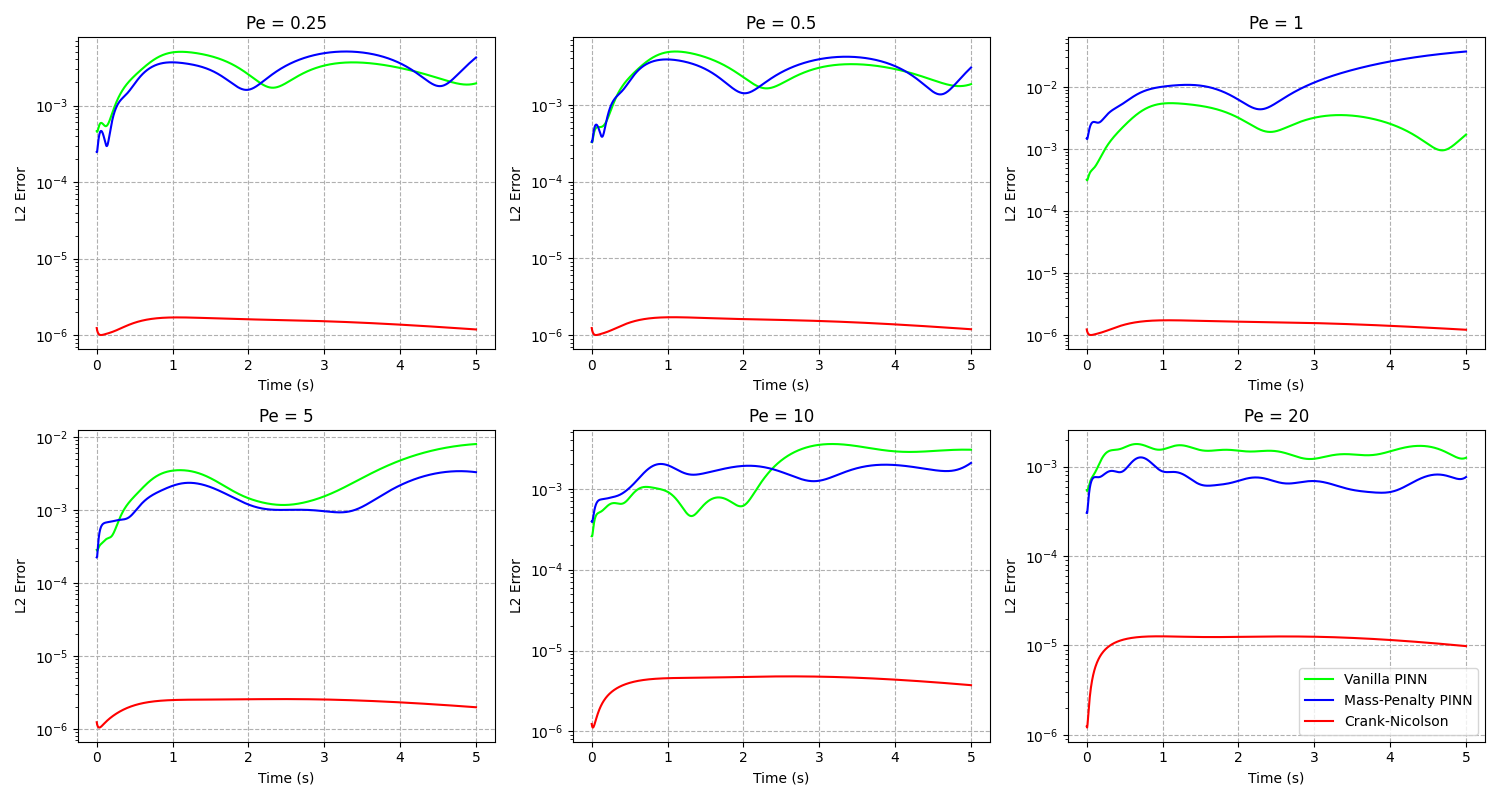}
\caption{Evolution of the relative $L_2$ error over time for different Peclet regimes ($T=5$\,s).}
\label{fig:l2_short}
\end{figure}

\begin{figure}[htbp]
\centering
\includegraphics[width=0.7\textwidth]{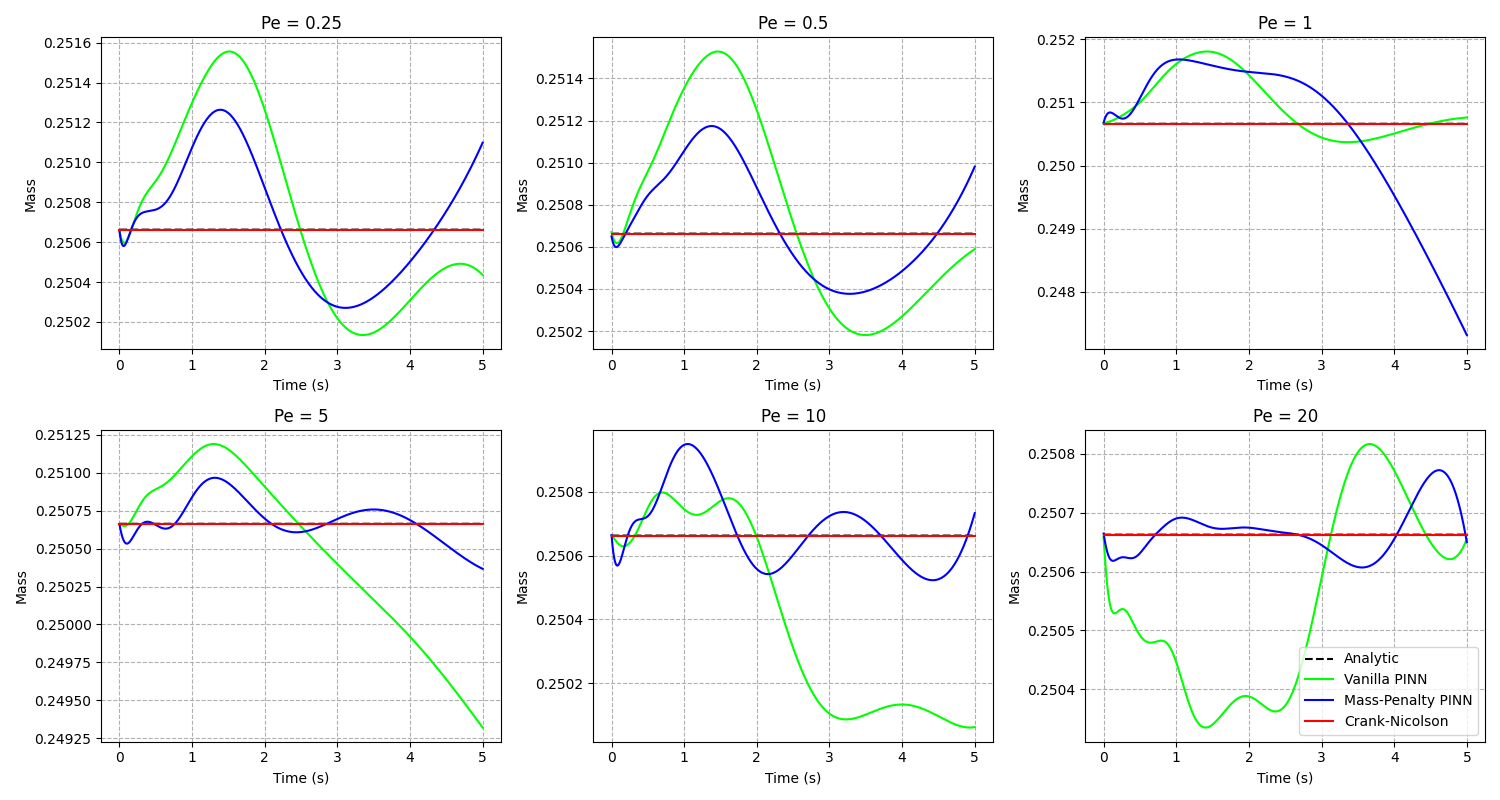}
\caption{Evolution of total mass over time for different Peclet regimes ($T=5$\,s).}
\label{fig:mass_short}
\end{figure}

For short-term simulations, both PINN models capture the concentration profiles adequately, yet their performance diverges depending on the flow regime. Figures \ref{fig:l2_short} and \ref{fig:mass_short} illustrate the evolution of relative $L_2$ error and total mass over time, respectively, for each Peclet number.

In diffusion-dominated regimes ($Pe \leq 1$), the Vanilla PINN slightly outperforms the Mass-Penalty PINN in numerical accuracy, because mass drift has not aggressively accumulated; however, the inclusion of the $\mathcal{L}_{\mathrm{mass}}$ constraint with $\lambda_{\mathrm{mass}}=10$ acts as a competing optimization objective that marginally hinders early convergence toward the exact local PDE solution. Conversely, in advection-dominated regimes ($Pe \geq 5$), the rapid displacement of mass exposes conservation errors earlier. Consequently, the Mass-Penalty PINN provides a distinct advantage here, yielding lower relative $L_2$ errors and strictly maintaining the global mass.

\subsection{Long-Term Simulation Performance ($T = 100$\,s)}

\begin{figure}[htbp]
\centering
\includegraphics[width=0.7\textwidth]{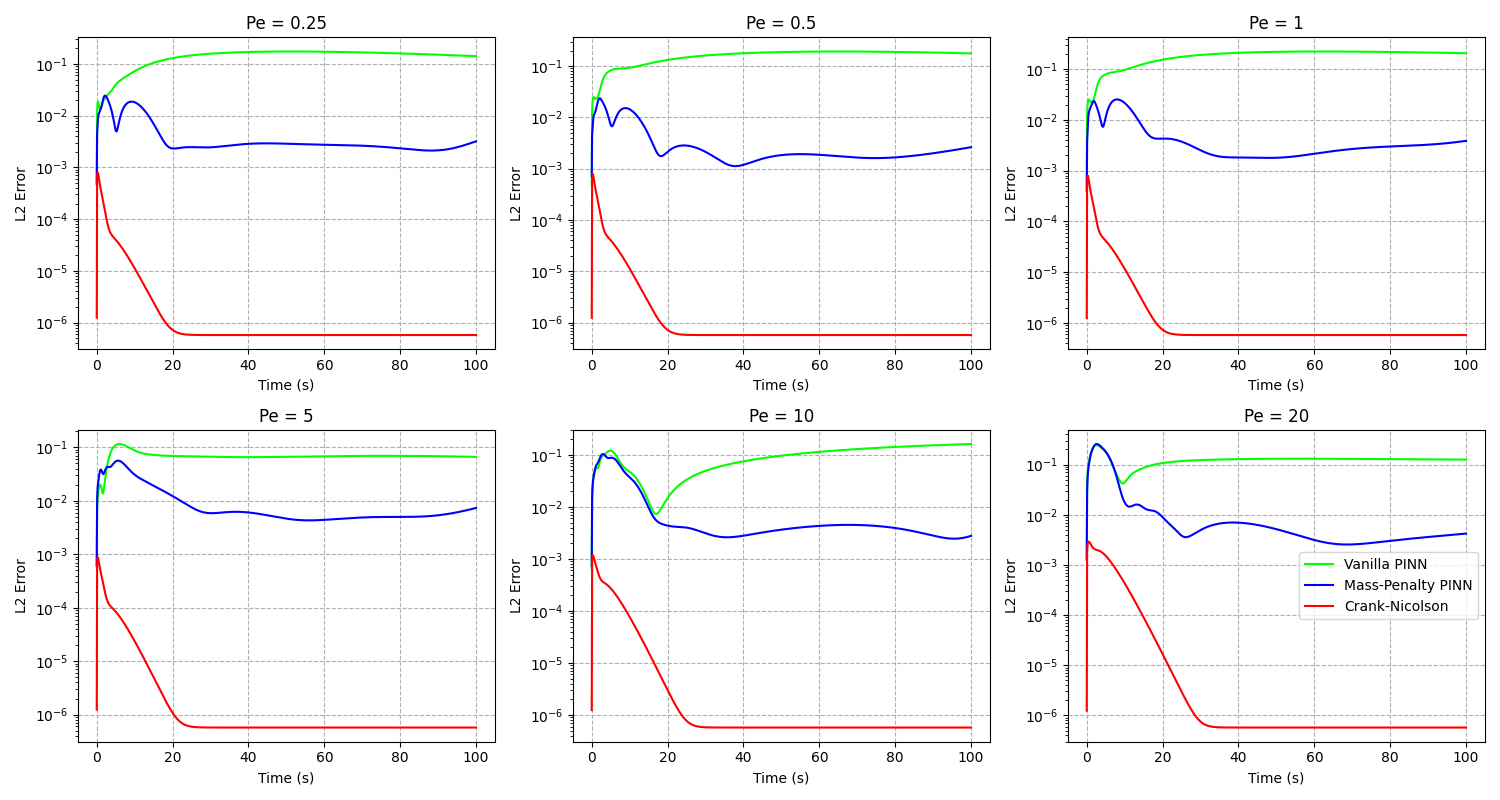}
\caption{Evolution of the relative $L_2$ error over time for different Peclet regimes ($T=100$\,s).}
\label{fig:l2_long}
\end{figure}

The long-term simulations reveal the cumulative impact of mass drift in Vanilla PINN, particularly in advection-dominated regimes. Figure \ref{fig:l2_long} shows the evolution of relative $L_2$ error over time for each Peclet number in the long-term simulation horizon.

\begin{figure}[htbp]
\centering
\includegraphics[width=0.7\textwidth]{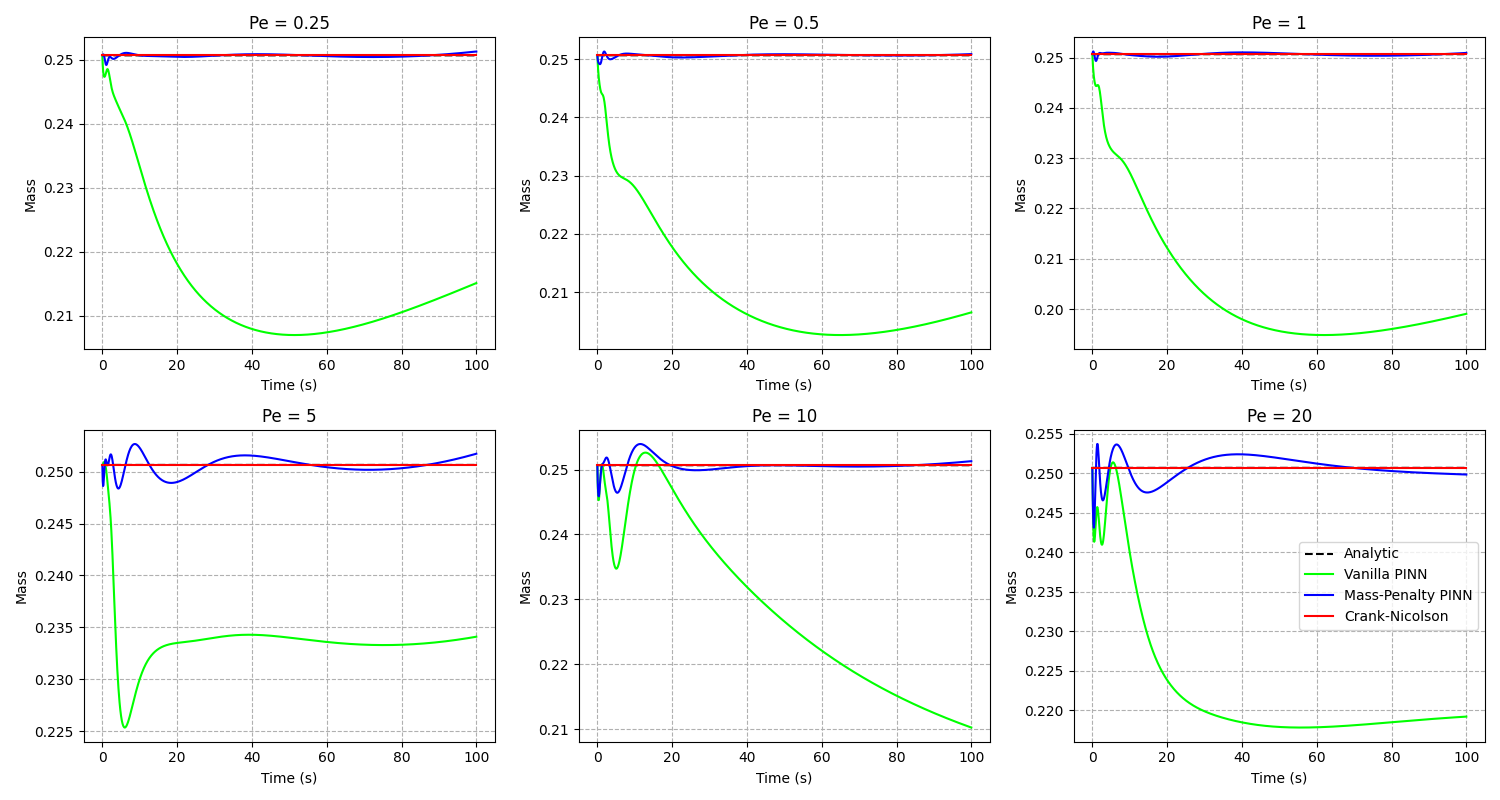}
\caption{Evolution of total mass over time for different Peclet regimes ($T=100$\,s).}
\label{fig:mass_long}
\end{figure}

Figure \ref{fig:mass_long} shows that under prolonged integration, the Vanilla PINN suffers from catastrophic mass drift. Specifically, the Vanilla formulation loses approximately 14\% to 21\% of its initial mass in diffusion-dominated states, and up to 16\% in advection-dominated states. Because the total mass dissipates fictitiously, the predicted concentration amplitudes severely underestimate the true physical states. The relative $L_2$ error of the Vanilla PINN scales proportionally with this mass loss, peaking at $2.06\times 10^{-1}$.

Integrating the mass-penalty constraint decisively eliminates this degradation. The Mass-Penalty PINN restricts mass deviation strictly between 0.08\% and 0.43\%. By forcibly anchoring the global mass, the final relative $L_2$ error stabilizes on the order of $10^{-3}$ across all flow regimes. The inclusion of this constraint improves the long-term accuracy by 9 to 67 times and reduces mass error by 15 to 215 times compared to the standard Vanilla approach.

\begin{figure}[htbp]
\centering
\includegraphics[width=0.4\textwidth]{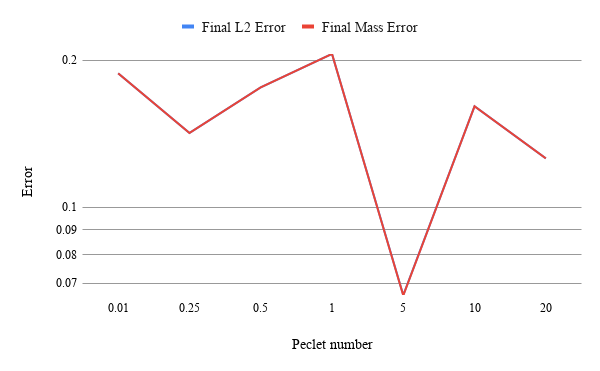}
\caption{Evolution of Relative $L_2$ error and mass error over time for different Peclet regimes ($T=100$\,s) using Vanilla PINN}
\label{fig:chart}
\end{figure}

In extended simulations where diffusion dominates, the analytical solution gradually approaches a homogeneous distribution. Under these conditions, the loss of mass leads the mean concentration predicted by the PINN to drift away from the true solution. Consequently, the relative $L_2$ error becomes nearly proportional to the relative mass error. This relationship is clearly reflected in Figure \ref{fig:chart}, where the final $L_2$ error of the Vanilla PINN is almost identical to its mass conservation error.

\subsection{Computational Efficiency}

A frequent criticism of strictly constrained deep learning is the computational overhead required to calculate dynamic spatial integrals during backpropagation.

\begin{table}[htbp]
\centering
\caption{Average training and inference time for each method.}
\label{tab:efficiency}
\begin{tabular}{@{}lcc@{}}
\toprule
Method & Avg. training time (s) & Avg. inference time (s) \\
\midrule
Vanilla      & $\approx 220.4$ & $\approx 0.123$ \\
Mass-Penalty & $\approx 232.4$ & $\approx 0.131$ \\
CN           & --              & $\approx 3.03$ \\
\bottomrule
\end{tabular}
\end{table}

Profiling results demonstrate that the Mass-Penalty PINN required an average training time of 232.4 seconds, compared to 220.4 seconds for the Vanilla PINN. This marginal 5.4\% increase in training cost is highly favorable when weighted against the exponential gain in long-term accuracy and physical validity. Furthermore, inference time remains entirely unaffected at $\approx 0.13$ seconds per evaluation. While the traditional Crank-Nicolson scheme remains highly efficient for single, forward simulation runs ($\approx 3.03$ seconds), the trained PINN evaluates solutions up to 25 times faster, making it advantageous for multi-query scenarios and inverse modeling tasks, where operator learning frameworks like DeepONet \cite{deeponet} or Fourier Neural Operators \cite{Li2020FourierNO} are also typically deployed.

\section{Discussion}

The results demonstrate that the long-term simulation of Vanilla PINN in conservative systems is primarily driven by accumulated mass drift rather than an inability to learn the spatial profile. The standard PINN loss function relies on pointwise PDE residuals. However, as shown by Wang et al. \cite{wang2021understand,wang2022when}, the optimization landscape of PINN is highly stiff, and the network often fails to satisfy global integral constraints because the gradients associated with global mass are outweighed by the local PDE residuals.

The mass-penalty constraint effectively introduces a global, zero-frequency anchor into the loss landscape. By penalizing deviations from the analytical mass, the optimizer is guided to preserve the integral invariant, thereby preventing spurious dissipation of the concentration field. It should be emphasized, however, that this remains a soft constraint. In highly advection-dominated regimes ($Pe \geq 10$), we observed minor fluctuations in local accuracy during intermediate training phases. This indicates that the soft penalty can occasionally interfere with the sharp gradients required to resolve steep advective fronts, a difficulty closely linked to the spectral bias of neural networks and the Kolmogorov n-width of advection-dominated manifolds \cite{MOJGANI2023115810}.

To further enhance local accuracy without sacrificing conservation, future work should explore hybrid approaches. Gradient-enhanced PINNs (gPINNs) \cite{yu2022gpinns} could be utilized to enforce conservation on the derivatives of the solution, while domain decomposition techniques like Extended PINNs (XPINNs) \cite{djagtap2020,SHUKLA2021110683} could isolate the stiff advective fronts into separate sub-networks. Additionally, transitioning from soft penalties to hard constraints where the network architecture is mathematically guaranteed to output a mass-conserving field remains an important direction for future research \cite{karlbauer2022composingpartialdifferentialequations,patsatzis2025gorinns}. Finally, although adaptive sampling strategies \cite{wu2022} were not applied in this baseline study, combining residual-based adaptive refinement with the mass penalty constraint could dynamically direct computational effort to regions with strong advective flux, thereby improving the robustness of the solution.

\section{Conclusion}

This study investigated the effect of enforcing mass conservation in Physics-Informed Neural Networks for the one-dimensional advection-diffusion equation. The results show that the standard Vanilla PINN is susceptible to the accumulation of non-physical mass drift, leading to significant accuracy degradation during long-term simulations. Incorporating a soft mass-penalty constraint effectively mitigates this issue, reducing the relative $L_2$ error and mass conservation error by up to 67 and 215 times, respectively, across the tested Peclet-number range, while increasing the training time by only 5.4\%. These findings indicate that the long-term deterioration of Vanilla PINNs is primarily driven by accumulated mass drift and demonstrate that enforcing global mass conservation substantially improves the physical consistency and reliability of PINN solutions. Although the Crank-Nicolson method remains the most accurate solver for this benchmark problem, the proposed Mass-Penalty PINN provides a promising mesh-free alternative for conservative transport problems, particularly in applications where the flexibility of PINNs is advantageous, such as inverse problems and parameter identification.

\end{document}